\documentclass[]{tJOT2e}

\usepackage{graphicx}
\usepackage{subfigure}
\usepackage{psfrag,epsfig,color}

\begin{document}
\doi{xxxx}
 \issn{00}
 \jvol{00} \jnum{00} \jyear{2014}

\markboth{M. Vo\ss kuhle {\sl et al}}{Collisions in suspensions -- comparing Navier-Stokes and synthetic turbulence}

\def\AP#1{{\color{blue}{#1}}}
\def\MW#1{{\color{red}{#1}}}
\def\EL#1{{\color{magenta}{#1}}}
\def\MV#1{{\color[cmyk]{0, 0.5, 1, 0}{#1}}}

\articletype{RESEARCH ARTICLE}

\title{
Collision rate for suspensions at large Stokes numbers -- comparing Navier-Stokes and synthetic turbulence
}

\author{Michel Vo\ss kuhle$^{\rm a}$, Alain Pumir$^{\rm a,d}$, Emmanuel L\'ev\^eque$^{\rm a,b}$, Michael Wilkinson$^{\rm c}$\\
\vspace{6pt}
$^{\rm a}$ Laboratoire de Physique, ENS de Lyon and CNRS, UMR5672, 46, all\'ee d'Italie, F-69007, Lyon, France.  \\
$^{\rm b}$ Laboratoire de M\'ecanique des Fluides et d'Acoustique, Ecole Centrale de Lyon and CNRS, UMR5509, 36, avenue Guy de Collonge,
F-69134, Lyon, France,   \\
$^{\rm c}$ Department of Mathematics and Statistics,
The Open University, Walton Hall, Milton Keynes, MK7 6AA, England \\
$^{\rm d}$ Max-Planck Institute for Dynamics and Self-Organisation,
D-37077, G\"ottingen, Germany.
\\ \vspace{6pt}
\received{February 2014}
}
\maketitle

\begin{abstract}
The use of simplified models of turbulent flows provides an
appealing possibility to study the collision rate of turbulent suspensions,
especially in conditions relevant to astrophysics, which require large
time scale separations.
To check the validity of such approaches, we
used a direct numerical simulation (DNS) velocity
field, which satisfies the Navier-Stokes equations (although
it neglects the effect of the suspended particles on the flow field),
and a kinematic simulation (KS) velocity field, which is a
random field designed so that its statistics are in accord
with the Kolmogorov theory for fully-developed turbulence.
In the limit where the effects of particle inertia (characterised
by the Stokes number) are negligible, the collision rates from the
two approaches agree. As the Stokes number ${\rm St}$ increases, however,
we show that the DNS collision rate exceeds the KS collision rate by orders of
magnitude. We propose an explanation for this phenomenon and explore its
consequences. We discuss the collision rate $R$ for particles in high Reynolds number
flows at large Stokes number, and present evidence that
$R\propto \sqrt{{\rm St}}$.
\end{abstract}

\section{Introduction}
\label{sec: 1}

A significant amount of work has been devoted in recent years
to the determination of the collision rate of small
particles suspended in a turbulent gas. This was primarily motivated by attempts
to understand rainfall from warm cumulus clouds \cite{Shaw03,GW13},
which depends upon
collisions of microscopic water droplets.
Another motivation comes from the efforts to
model planet formation \cite{Safranov:69} involving aggregation of
dust grains in turbulent circumstellar accretion discs. Collisions are also
important in determing the properties of particle-laden turbulent flows
such as snow avalanches and sandstorms \cite{Elghobashi:94}.

The calculation of collision rates is a complex problem.
It was appreciated early on that
collision rates strongly depend on
the inertia of the suspended particles, an effect which can be
characterized by the Stokes number:
\begin{equation}
{\rm St}=\frac{\tau_{\rm p}}{\tau_{\rm K}}
\label{eq: 1}
\end{equation}
where $\tau_{\rm p}$ is the timescale for particle
velocities to relax to the ambient flow, and $\tau_{\rm K}$ is the
Kolmogorov time scale : $\tau_{\rm K} \equiv (\nu/\epsilon)^{1/2}$, where
$\epsilon$ is the rate of dissipation per unit mass and $\nu$ the kinematic viscosity
of the fluid.
For ${\rm St}\ll 1$, particles are advected by the fluid, and collisions are
the result of shear.
The small ${\rm St}$ limit is particularly relevant to
the problem of rain initiation~\cite{Shaw03}.
When ${\rm St}\gg 1$, the inertia of the particles allows
them to move relative to the surrounding fluid.
Two very different physical situations may occur when ${\rm St}$ is large.
Turbulent flows are characterized by a large range of spatial and temporal
scales. The time scales of eddies of scale $l$ is of the order
of $t_0 \sim (l^2/\epsilon)^{1/3}$. The ratio between the time scale of
the largest eddies,~$T_L$, and the smallest eddies, $\tau_{\rm K}$,
is known to be $T_L/\tau_{\rm K}
\approx 0.08 {\rm Re}_\lambda$~\cite{SY:2011}, where ${\rm Re}_\lambda$ is the
Taylor microscale Reynolds number. In very turbulent flows, relevant
to astrophysical problems,
\begin{equation}
1 \ll {\rm St} \lesssim T_L/\tau_{\rm K}
\label{eq:astro_cond}
\end{equation}
a condition expressing that the largest eddies have a longer time scale
than $\tau_{\rm p}$. The other case,
$\tau_{\rm p} > T_L$ is likely to probe a very different physical
regime.
Reliably determining the collision rate with the condition
\eqref{eq:astro_cond}
requires simulation of flows with a very large range of
spatial and temporal scales,
a very demanding endeavour.

The most realistic evaluation of the collision rate is obtained
by using direct numerical simulation (DNS), where the velocity field satisfies
the Navier-Stokes equations (to within numerical errors).
There is a substantial literature on DNS collision rates, some notable
contributions
are \cite{Sundaram:96,Sundaram:97,Wang:00,RC:00,Rosa:13,FP07}.
It is however notoriously difficult to perform well-controlled DNS
studies at very large values of the Reynolds number, ${\rm Re}$, because
the grid size,
which determines the ratio of the integral scale to the Kolmogorov scale,
is limited to numbers of order $(10^3)^3$. For most purposes, DNS calculations
with this resolution provide a good approximation to the ${\rm Re}\to \infty$
limiting behaviour. The exception, however, occurs when we consider
processes which probe long timescales in the flow, as
effectively imposed by the condition~\eqref{eq:astro_cond}.

The alternative approach to DNS, consisting in using a randomly fluctuating
velocity field whose properties mimic those of turbulent flows,
has two potential advantages.
A potentially attractive
feature is that if the statistics of the velocity field are known, it may be possible
to compare with analytical theories.
The principal benefit is that a random vector field requires fewer numerical
operations and allows
simulations with
a larger range of scales,
containing long-lived eddies.
One especially refined version of the random velocity
field approach is termed kinematic simulation (KS)~\cite{Fung:92}.
The KS velocity field contains
several parameters,
which introduces uncertainties into the collision rates.
Few studies use KS to determine collision rates~\cite{DP09,Ijzermans10}.
We are not aware of earlier work directly comparing DNS and KS approaches.

An enticing approach could consist in combining the use of both DNS and KS velocity fields.
Using the collision rate determined from DNS at intermediate
Stokes numbers, ${\rm St } \lesssim 1$,
one could adjust the parameters of the KS velocity field to match the DNS collision
rate. Then at large Stokes numbers, the collision rate could be
obtained from KS simulations.

We attempted to use this approach to investigate
collision rates at large Stokes numbers. Unexpectedly, we found that the
approach is not viable, because at Stokes number ${\rm St} \sim 1$, the
DNS collision rates
exceed the KS rates by a large factor, which cannot be reduced
by adjusting parameters of the KS simulation.
Figure \ref{fig: 1} illustrates the comparison between
collision rates evaluated using the two approaches: they
agree only at very
small Stokes number.

In this paper we document this observation, propose an explanation for the
discrepancy, and discuss the problem of making a quantitatively
accurate estimate for the collision rate at large Stokes numbers.
We also remark that there is a similar discrepancy between DNS and KS
simulations in the clustering behaviour of particles.

\section{Theoretical background}

In this section, we briefly review the available understanding of the
collision rate in turbulent suspensions. The main result, \eqref{eq: 5},
expresses the collision rate as a sum of two contributions, originating from
two different effects induced by the particle inertia.
In the limit of large Stokes numbers,
under the condition \eqref{eq:astro_cond}, the limiting behaviour of the collision rate
is predicted to be given by a simple analytic expression, \eqref{eq: 6}.

For a suspension of spherical particles of radius $a$, the rate of collision of a given
particle with any other particle may be written
\begin{equation}
\label{eq: 2}
R=2\pi n_0 a^2 \langle w\rangle
\end{equation}
where $n_0$ is the number density of particles, and
$\langle w\rangle$ is a suitably defined average of the
relative velocity of two particles when their separation is equal
to $2a$~\cite{Sundaram:97,Wang:00}.
The collision rates have been investigated by a succession of authors using
simulations where particles move independently
under the simplified Gatignol/Maxey-Riley equations of motion \cite{Gat83,MaxRil83}.
\begin{equation}
\dot{\mbox{\boldmath$r$}}=\mbox{\boldmath$v$}
\ ,\ \ \
\dot{\mbox{\boldmath$v$}}=\frac{1}{\tau_{\rm p}}[\mbox{\boldmath$u$}(\mbox{\boldmath$r$},t)-\mbox{\boldmath$v$}]
\label{eq: 3}
\end{equation}
where $\tau_{\rm p}=\frac{2}{9}\frac{a^2}{\nu}\frac{\rho_{\rm p}}{\rho_{\rm f}}$
is the particle relaxation time, determined from Stokes formula for the
drag on a moving sphere, $\rho_{\rm f}$ and $\rho_{\rm p}$ being
respectively the fluid and particle densities. These equations of
motion are valid in the limit
where the suspended particles are very small and very dense:
$\rho_{\rm p}/\rho_{\rm f} \gg 1$. Most works have defined
the collision rate as being the rate for the separation
between non-interacting particles decreasing past $2a$.
In practice, this \lq ghost collision approximation' may
overestimate the collision rate, because collisions may be
inhibited by lubrication effects, and because multiple collisions
should not be counted if particles adhere, coalesce, or react
on contact~\cite{Vosskuhle:13}.
In this work we use the ghost collision approximation,
because our objective is to isolate the effects of turbulence
from other phenomena.

The early work of Saffman and Turner
\cite{ST56} considers the case where the Stokes number is small, so that the suspended
particles are advected with the flow. In this limit collisions
occur due to shearing motion,
so that $\langle w\rangle\sim a/\tau_{\rm K}$.
Saffman and Turner
showed
that if the \lq ghost collision' criterion
is used as the definition of the collision rate, then
for ${\rm St}\ll 1$
the rate
of collision of a given
spherical particle of radius $a$ with any other particle is
\begin{equation}
\label{eq: 4}
R_{\rm ST}=\sqrt{\frac{8 \pi}{15} } \frac{n_0 (2 a)^3}{\tau_{\rm K}}
\ .
\end{equation}
This expression is exact in the limit as ${\rm St}\to 0$, and
in this paper equation (\ref{eq: 4}) will be used as a benchmark for comparing
collision rates in turbulent flow.

When the Stokes number is not small, the inertia of the suspended particles allows them
to follow trajectories which differ from fluid streamlines. This brings two additional
effects into play, which can substantially increase the collision rate.
The first mechanism which has been proposed to
enhance the collision rate \cite{Sundaram:96} is the tendency for
inertial particles in turbulent flows to exhibit a clustering effect, known
as preferential concentration \cite{Max87}.
This has the effect of introducing an
additional factor in equation (\ref{eq: 4}): the collision rate is multiplied
by $g(2a)$, where $g(r)$ is the radial distribution function describing the clustering effect.
The other mechanism for enhancing the collision rate is that the particle velocity can become
multivalued, due to the formation of caustic folds in the phase-space of the
suspended particles \cite{WM05}, and this effect can lead
to an enhancement of the collision rate \cite{WMB06}. The same mechanism has
also been described as a \lq sling effect', where particles collide because
they are centrifuged out of vortices \cite{FFS02}.

These inertial effects are included by adopting the
following model for the collision rate
\begin{equation}
\label{eq: 5}
R=R_{\rm ST}\,g(2a)+n_0a^2u_{\rm K}\,f({\rm St},{\rm Re})
\end{equation}
where
$f$ is a function
of the Stokes number and the Reynolds number.
The factor $g(2a)$ accounts for the
enhancement of the advective collision mechanism by the preferential
concentration
effect, as has been explained above.
The other term describes the collisions of particles whose
velocities differ significantly from the flow velocity,
so that their relative velocity is proportional
to $u_{\rm K}$, i.e., does not vanish in the limit of very small particle
separation, and therefore, for small, dense particles greatly exceeds $a/\tau_{\rm K}$.
The function $f({\rm St},{\rm Re})$ approaches zero rapidly as ${\rm St}\to 0$, so that
(\ref{eq: 4}) is valid in this limit. The theoretical basis for equation
(\ref{eq: 5}) was set out in \cite{FFS02,WMB06,Gustavsson:11}, and
recent DNS simulations at moderate Stokes numbers (${\rm St} \le 5$)
\cite{Vosskuhle:14}
lend strong support to its validity. In particular, simulations do show a
marked increase in the collision rate when ${\rm St}$ is of order unity,
and the DNS simulations are consistent with theoretical expectations,
at least at moderate Stokes numbers.

As explained in the introduction, the collision rate at very
large Stokes numbers is an important
issue for models of planet formation, and may also be significant
for models of particle-laden geophysical flows. Different lines of
argument, explicitly using the condition \eqref{eq:astro_cond}
\cite{Vol+80,Mehlig:07} indicate that
\begin{equation}
\label{eq: 6}
f({\rm St},\infty)\sim K{\rm St}^{1/2}
\end{equation}
as ${\rm St}\to \infty$.
Providing numerical evidence to establish the validity of \eqref{eq: 6},
and to determine the coefficient $K$ would lead to an explicit and
reliable parametrisation of the collision rate in the important limit
of large Stokes numbers.

However, DNS studies are limited to quite small
values of ${\rm St}$, and it has not been possible to see conclusive
evidence that $f({\rm St},\infty)\propto \sqrt{{\rm St}}$, or to determine the
coefficient $K$ in (\ref{eq: 6}).
The difficulty in investigating the validity of \eqref{eq: 6} using
directly DNS is one the main reasons to look for alternative numerical
methods, based on simplified KS flow models.

\section{Numerical studies comparing DNS and KS}
\label{sec: 3}

It appears to be possible to study the collision rate at large
Stokes number by replacing the DNS velocity field with a
KS velocity field, because generating the synthetic field requires fewer
numerical operations, and allows investigations of larger
systems.

The details
of how we implemented this KS velocity field model are described
in detail in~\cite{DP09}, and are summarised briefly here.
The velocity field is
\begin{equation}
\label{eq: 7}
\mbox{\boldmath$u$}(\mbox{\boldmath$r$},t)=\sum_{n=1}^{\cal N}
\Bigl(
\mbox{\boldmath$a$}_n \cos(\mbox{\boldmath$k$}_n\cdot \mbox{\boldmath$r$}+\omega_n t)
+ \mbox{\boldmath$b$}_n \sin(\mbox{\boldmath$k$}_n\cdot \mbox{\boldmath$r$}+\omega_n t)
\Bigr)
\end{equation}
where $\mbox{\boldmath$k$}_n\cdot \mbox{\boldmath$a$}_n=\mbox{\boldmath$k$}_n\cdot \mbox{\boldmath$b$}_n=0$
(to ensure that the field is incompressible) and the vectors $\mbox{\boldmath$a$}_n$ and
$\mbox{\boldmath$b$}_n$ are chosen at random, with the variances of their
elements prescribed so that the velocity field has a spectral density which is
in agreement with the Kolmogorov law: $E(\mbox{\boldmath$k$})=E_0|\mbox{\boldmath$k$}|^{-5/3}$.
The choice of the upper cutoff for the wavevectors determines the Kolmogorov length $\eta$,
and the coefficient $E_0$ the determines the rate of dissipation $\epsilon$.
The frequencies $\omega_n$ are determined by the relation between lengthscale and
eddy turnover time which is implied by Kolmogorov's dimensional analysis:
\begin{equation}
\label{eq: 8}
\omega_n=\lambda \sqrt{|\mbox{\boldmath$k$}_n|^3E(\mbox{\boldmath$k$})}
\end{equation}
where $\lambda$ is a dimensionless coefficient, which determines
the phase velocity of the modes.
The value of the coefficient $\lambda $ is not predicted
by the Kolmogorov theory.
It determines the Kubo number of the flow: ${\rm Ku}=u_0\tau/\xi$,
where $u_0$, $\tau$ and $\xi$ are, respectively, velocity, length and
timescales associated with the smallest eddies in the flow. The Kolmogorov
theory identifies these quantities as being
proportional to $u_{\rm K}$, $\tau_{\rm K}$ and $\eta$, respectively.
Dimensional arguments
then establish that ${\rm Ku}$ is of order unity,
but the precise value of $\lambda$ is unknown.

These considerations lead to the following proposal: that the KS simulation
can be \lq calibrated' against DNS simulations to determine the appropriate value
for $\lambda$: that is, the value of $\lambda$ should be chosen to make the KS collision
rate match the DNS rate as closely as possible at small values of the Stokes number,
where reliable DNS data can be determined. The collision rate could then be extrapolated
to high values of the Stokes number using KS simulations.

\begin{figure}
\begin{center}
\subfigure[]{
\includegraphics[width=0.45\textwidth]{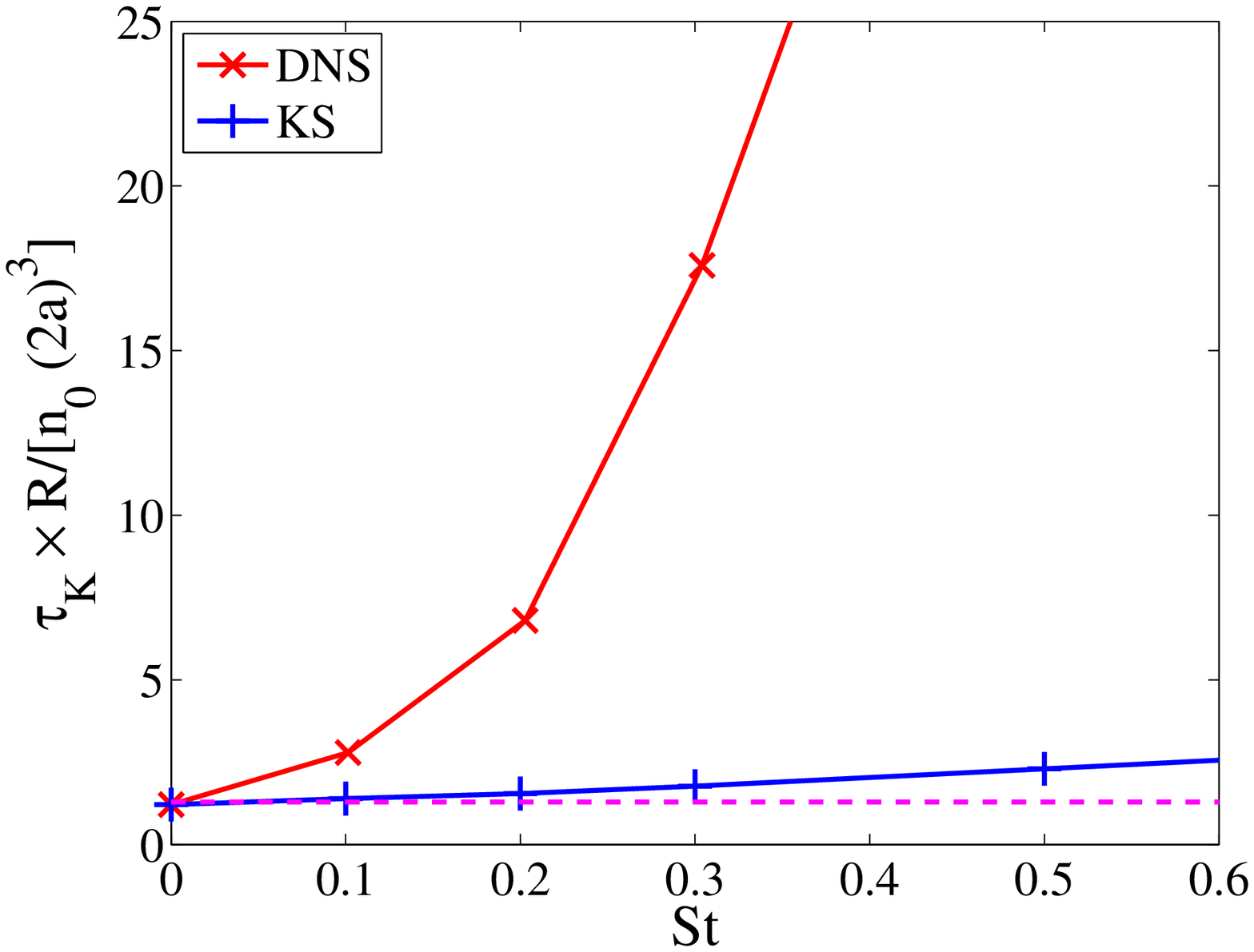}
}
\subfigure[]{
\includegraphics[width=0.45\textwidth]{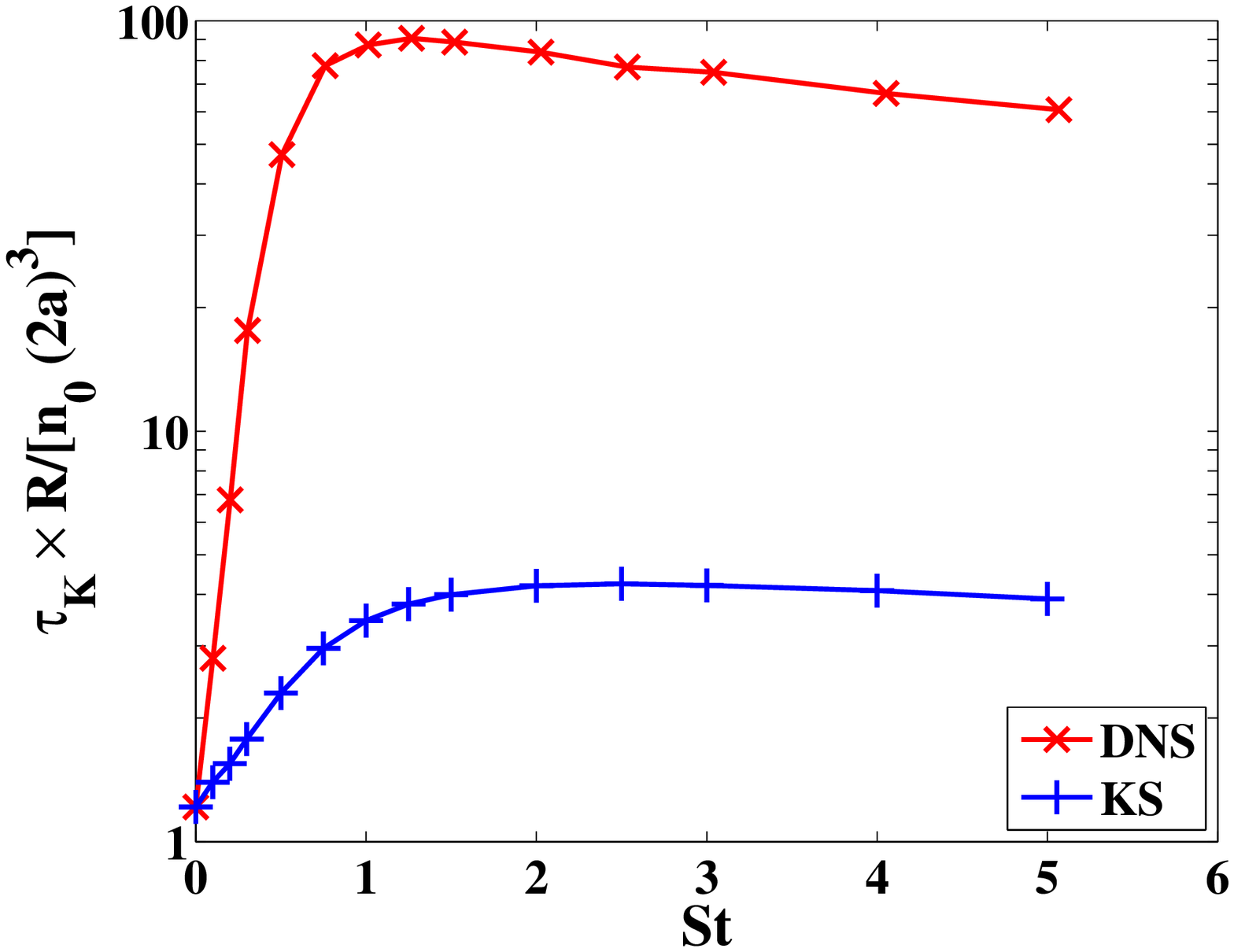}
}
\caption{\label{fig: 1}
(Colour online). Comparison of collision rates evaluated using DNS
and KS velocity fields as a function of the Stokes number: ({\bf a}),
linear scale, shows that the collision rates agree at small Stokes number,
({\bf b}), logarithmic scale. The predictions of DNS and KS agree only
in the limit ${\rm St}  \rightarrow 0$, and coincide with the
prediction of~\cite{ST56}, shown by the dotted line. At values of the
Stokes number
larger than ${\rm St } \gtrsim 0.3$, the collision rate $R$
determined with the KS flow is smaller by more than one order of magnitude
compared to the collision rate obtained in DNS.
}
\end{center}
\end{figure}

We tested this approach, using the methods described in
\cite{Vosskuhle:14,Vosskuhle:13} to compute the DNS collision rate.
The procedure was found to fail, however: the DNS collision
rate rapidly becomes much larger than the KS collision rate as ${\rm St}$
increases. The two collision rates are compared in Fig.~\ref{fig: 1} using
both  linear and logarithmic scales.
The discrepancy is so marked that no
adjustment of $\lambda$ can make the two collision rates agree exactly,
except in the limit as ${\rm St}=0$.
This is illustrated in table \ref{table:1}, which shows the dependence of
the normalized collision rate, $R \tau_{\rm K}/(2a)^3$ on $\lambda$,
at the value ${\rm St} = 1.5$, and for a moderate scale ratio $L/\eta = 64$.
The normalized collision rate is found to be a decreasing
function of $\lambda$. However, when $\lambda $
decreases from $\lambda = 0.5$, which is the value taken in most simulations,
to $\lambda = 0.01$, the normalized collision rate, $R \tau_{\rm K}/(2a)^3$
increases by less than a factor of $2$ at ${\rm St } = 1.5$. This is
insufficient to correct
for the difference, clearly seen in Fig.~\ref{fig: 1}, between the DNS and KS results.
Increasing $\lambda $ to the value $2$,
on the other hand decreases the collision rate by a factor $\sim 1.5$.
We checked that these conclusions, based on the data shown in
table \ref{table:1}, extend to the entire range of Stokes numbers investigated
here.

\begin{table}[h]
\tbl{%
The dependence of the collision rate, normalized by $\tau_{\rm K}/n_0(2a)^3$,
on the parameter $\lambda$ (defined by \eqref{eq: 8}), at a fixed
value of the Stokes number
(${\rm St} = 1.5$) and of the
ratio of scales for $L/\eta = 64$.
}{%
\begin{tabular}{@{}lccccc}
\toprule
$\lambda$
& $0.01$
& $0.1$
& $0.5$
& $1.0$
& $2.0$
\\\colrule
$R \tau_{\rm K} /n_0 (2a)^3$
& $6.3$
& $5.7$
& $4.0$
& $3.1$
& $2.6$
\\\botrule
\end{tabular}
}
\label{table:1}
\end{table}

The DNS and KS collision rates agree exactly in the limit ${\rm St}\to 0$
because $R_{\rm ST}$ can be expressed in terms of the expectation value
of the trace of the square of the strain-rate matrix. Because this
quantity depends only upon the spatial derivatives of the velocity field,
and not upon its kinematics, the KS collision rate for ${\rm St}\to 0$
is asymptotic to $R_{\rm ST}$, independent of the value of $\lambda$.

\begin{figure}
\begin{center}
\subfigure[]{
\includegraphics[width=0.45\textwidth]{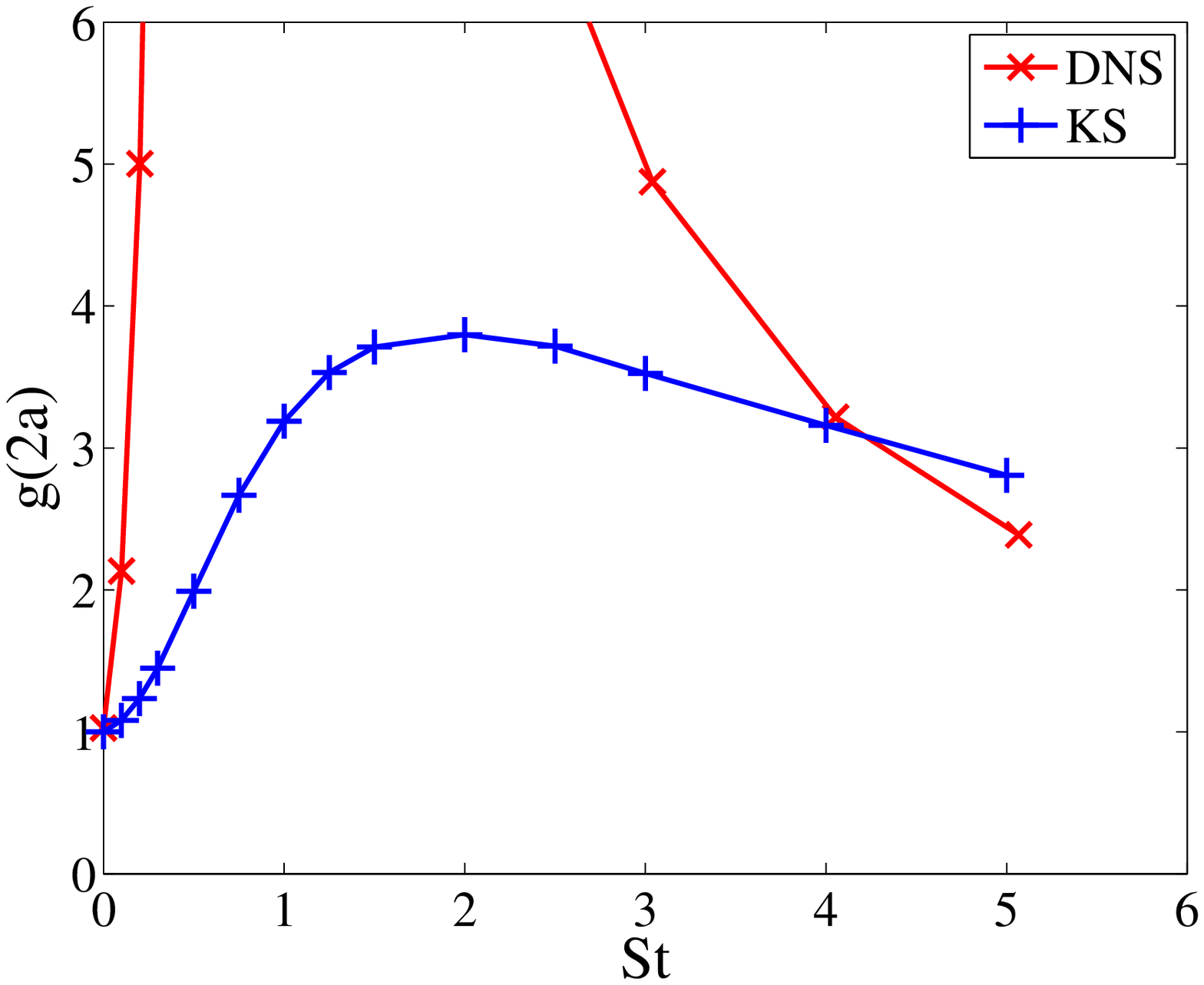}
}
\subfigure[]{
\includegraphics[width=0.45\textwidth]{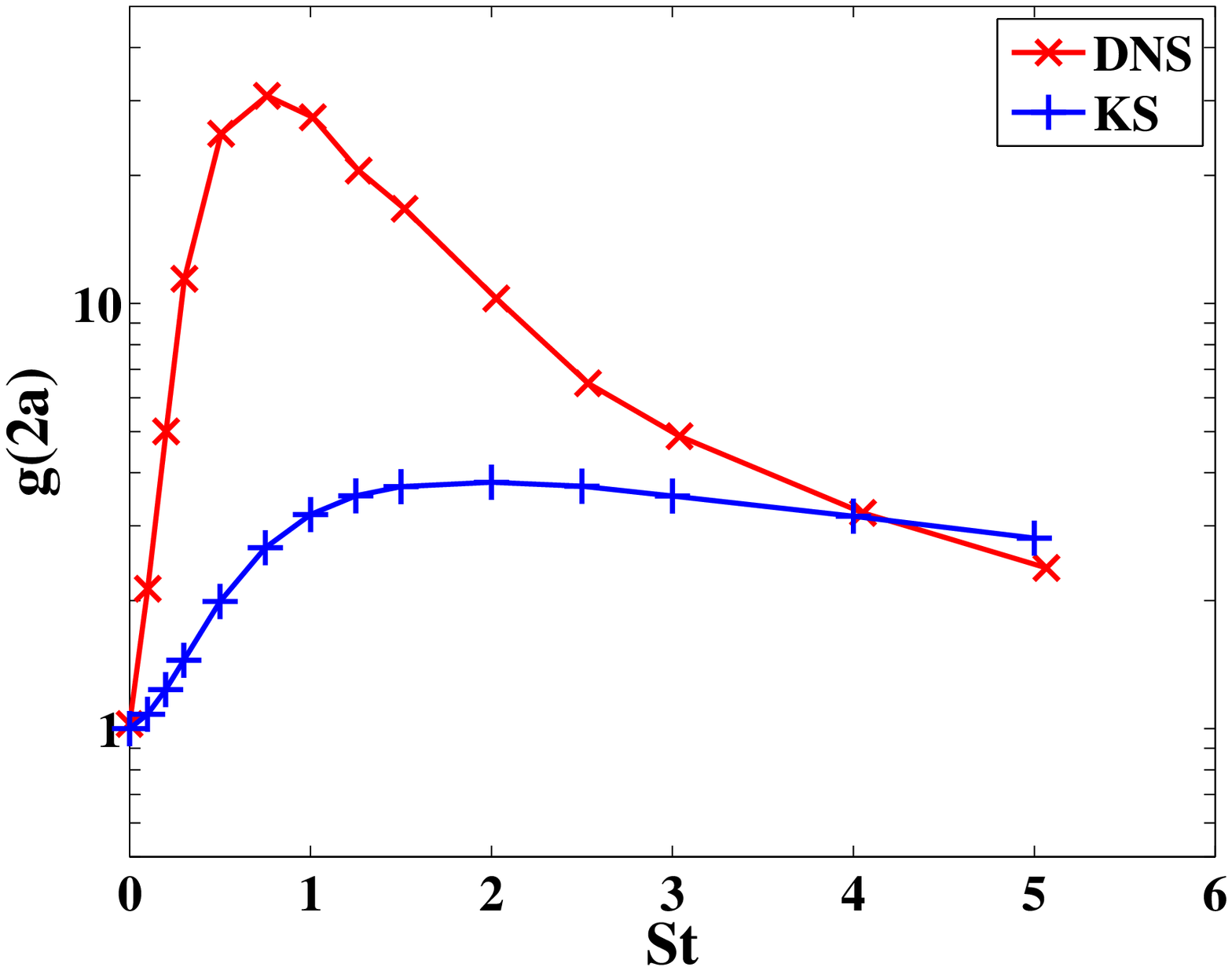}
}
\caption{\label{fig: 2}
(Colour online). Comparison of the preferential concentration, measured
 by the pair correlation function $g(2a)$, as a function of the Stokes number:
({\bf a}),
linear scale, shows that the preferential concentration effect is not present
at small Stokes number,({\bf b}), logarithmic scale, shows that the preferential
concentration in the KS flow field is too low
by orders of magnitude at larger Stokes number.}
\end{center}
\end{figure}

We also compared the preferential concentration effect for the two
velocity fields, finding that the preferential concentration is
much weaker in KS velocity fields, as compared to the equivalent DNS field.
This is illustrated in figure \ref{fig: 2}.

\section{Discussion}
\label{sec: 4}

Our numerical studies indicate that seeking to extrapolate
numerical estimates of collision rates to large Stokes numbers
using KS velocity fields is not a viable program, because the
KS data become very inaccurate as the Stokes number ${\rm St}$
increases past a value of the order ${\rm St } \gtrsim 0.1$, see
Fig.~\ref{fig: 1}.

This observation is also significant for theoretical discussions
of collision rates, which consider the rate of collision processes
in randomly defined velocity fields. The KS velocity field is just
a specific example of a random velocity field. Our work shows
that caution must be applied when interpreting the results of
these calculations.

We should consider why the KS model performs so badly.
The following argument indicates the probable
source of the discrepancy. In the realistic DNS model,
both the particles and the smaller eddies are
swept along by the larger eddies. In the KS simulations,
however, the particles are swept but the small \lq eddies' don't
move. This means that the fluctuations caused
by the small eddies (those which create the relative
velocities) have a reduced correlation time (when
viewed by the particles) and are therefore less effective.
The caustics~\cite{WM05}, inducing a multivalued velocity distribution
of the particles~\cite{FFS02}, and constituting
the predominant mechanism for
increasing the collision rate~\cite{Vosskuhle:14}, are generated at a
slower rate in the case of the KS velocity field, with a consequent
reduction of the collision rate.

Random flow field models which use single-scale velocity
fields have been very successful in explaining the qualitative
features of the role of caustics in enhancing collision rates
\cite{WMB06,Gus+13}. They have also been shown to be able to
give quantitatively correct results in describing preferential
concentration \cite{Wil+07}. Our results indicate that attempts
to improve upon the predictive ability of random flow models
for turbulence by incorporating the multi-scale aspect of the
flow seem to be unsuccessful. The reason for this failure can be attributed
to the lack of sweeping of the small eddies by the large eddies in the KS
model. This very important difference of the KS flow, in comparison to
DNS has been shown to lead to different predictions
~\cite{Thomson:05}. In particular, it has been qualitatively noticed that
preferential concentration is reduced in KS \cite{Chen:06}.
We find that even in the range of Stokes number ${\rm St} \gtrsim 1$, where
the enhancement of the collision rate is not so much due to preferential
concentration, but rather to the caustics effect~\cite{Vosskuhle:14}, the
KS model seriously underestimates the  collision rates.

\begin{figure}
\begin{center}
\includegraphics[width=0.65\textwidth]{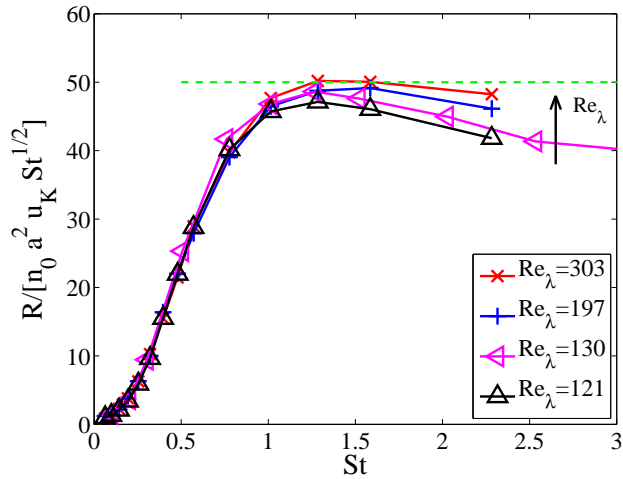}
\caption{\label{fig: 3}
(Colour online). Plot of the collision rate $R$ divided by $n_0 a^2 u_{\rm K}\sqrt{{\rm St}}$,
as a function of ${\rm St}$. The curves appear to approach a plateau at large ${\rm St}$ as
 the Reynolds number increases, consistent with equation (\ref{eq: 9}). The data for
 ${\rm Re}_\lambda=130$ is from \cite{Vosskuhle:14}, the other data is re-plotted from \cite{Rosa:13}.}
\end{center}
\end{figure}

The evaluation of the collision rate in turbulence at large Stokes and Reynolds
remains a significant open problem. The results presented here indicate that
only DNS evaluations should be considered reliable. The best
approach is to utilise DNS data at the largest Stokes number for which
a DNS collision rate has been observed to be substantially independent of the Reynolds
number. We used data from a high resolution
study of collisions in turbulent flows by Rosa {\sl et al.}, \cite{Rosa:13}, together
with our own data from \cite{Vosskuhle:14}. In accord with arguments in \cite{Vol+80,Mehlig:07}
and with equations (\ref{eq: 5}) and (\ref{eq: 6}) above, we fitted a collision
rate proportional to $\sqrt{{\rm St}}$. The data plotted in figure \ref{fig: 3} indicate
that the collision rate for turbulence with very high Reynolds numbers is
\begin{equation}
\label{eq: 9}
R\approx K n_0a^2u_{\rm K}\sqrt{{\rm St}}
\ , \ \ \ \
K=50
\ .
\end{equation}
for large values of the Stokes number, with the plateau being reached
by ${\rm St}\approx 1.2$.
Note that, while the curves show a decrease for
larger values of ${\rm St}$, they appear to approach a plateau as the Reynolds
number increases.
The position of the plateau and therefore the exact value of the constant $K$ seem to depend (slightly) on the Reynolds number.
But until
simulations at higher Reynolds number become available,
equation~\eqref{eq: 9}
is
the best available estimate for the collision rate of monodisperse
spherical particles at high Stokes number.

The authors acknowledge informative discussions with B. Mehlig,
K. Gustavsson and L. Collins. AP and EL have been supported by the grant from
A.N.R. \lq TEC 2'. Computations were performed at the PSMN computing center
at the Ecole Normale Sup\'erieure de Lyon.
MW and AP were supported by the EU COST action MP0806 \lq Particles in Turbulence'.
AP acknowledges the support from the Alexander-von-Humboldt foundation.

\end{document}